# The Mutual Orbit, Mass, and Density of the Large Transneptunian Binary System Varda and Ilmarë

W.M. Grundy[a], S.B. Porter[b], S.D. Benecchi[c], H.G. Roe[a], K.S. Noll[d], C.A. Trujillo[e], A. Thirouin[a], J.A. Stansberry[f], E. Barker[f], and H.F. Levison[b]

a. Lowell Observatory, 1400 Mars Hill Rd., Flagstaff AZ 86001.
b. Southwest Research Institute, 1050 Walnut St. #300, Boulder CO 80302.
c. Planetary Science Institute, 1700 E. Fort Lowell Suite 106, Tucson AZ 85719.
d. NASA Goddard Space Flight Center, Greenbelt MD 20771.
e. Gemini Observatory, 670 N. A'ohoku Place, Hilo HI 96720.
f. Space Telescope Science Institute, 3700 San Martin Dr., Baltimore MD 21218.

—— In press in *Icarus* ——

## Abstract

From observations by the Hubble Space Telescope, Keck II Telescope, and Gemini North Telescope, we have determined the mutual orbit of the large transneptunian object (174567) Varda and its satellite Ilmarë. These two objects orbit one another in a highly inclined, circular or near-circular orbit with a period of 5.75 days and a semimajor axis of 4810 km. This orbit reveals the system mass to be $(2.664 \pm 0.064) \times 10^{20}$ kg, slightly greater than the mass of the second most massive main-belt asteroid (4) Vesta. The dynamical mass can in turn be combined with estimates of the surface area of the system from Herschel Space Telescope thermal observations to estimate a bulk density of $1.24^{+0.50}_{-0.35}$ g cm$^{-3}$. Varda and Ilmarë both have colors similar to the combined colors of the system, $B$–$V = 0.886 \pm 0.025$ and $V$–$I = 1.156 \pm 0.029$.

## 1. Introduction

The Kuiper belt, an enormous region beyond Neptune's orbit, is populated by icy planetesimals left over from the formation of the giant planets. Roughly a quarter million trans-neptunian objects (TNOs) larger than 100 km are estimated to reside in this region (e.g., Petit et al. 2011; Gladman et al. 2012). The largest few are planet-sized bodies such as Pluto, Eris, and Makemake, massive enough to retain atmospheres in vapor pressure equilibrium with volatile surface ices (e.g., Elliot et al. 1989; Schaller and Brown 2007). Numerous slightly smaller TNOs are large enough to relax into spherical shapes, and could have had some degree of geological activity in the past, or even still be active today (e.g., Desch et al. 2009; Malamud and Prialnik 2015). The scope for comparative planetological studies of these bodies is tremendous, since they come in a range of sizes, occupy a variety of heliocentric orbits, and likely accreted at different heliocentric distances within the protoplanetary nebula, drawing on chemically distinct compositional reservoirs of solid materials. Despite this scientific promise, much has still to be learned about these bodies. For the best characterized among them, their sizes, albedos, surface colors and compositions, masses, densities, and spin states are known, but prior to spacecraft exploration, knowledge of surface geology, interior structure, and bulk composition will remain largely conjectural. For many, even the most fundamental parameters have not yet been determined. Efforts to measure them continue.

Transneptunian object (174567) Varda, the subject of this paper, was discovered in 2003 by J.A. Larsen et al. at Steward Observatory's 0.9 m Spacewatch telescope on Kitt Peak (Larsen et al. 2007), and initially assigned a provisional designation of 2003 $MW_{12}$. Varda orbits the Sun on an inclined ($i_\odot = 21°$) and eccentric ($e_\odot = 0.15$) orbit that would be classified as "Scattered Extended" in the Deep Ecliptic Survey system (e.g., Elliot et al. 2005; http://www.boulder.swri.edu/~buie/kbo/desclass.html) and "Classical" in the Gladman et al. (2008) system. The high inclination excludes membership of the dynamically cold core of the Classical belt, where, at least among the brighter objects, binary rates are especially high and colors are especially red (e.g., Noll et al. 2008; Gulbis et al. 2010; Petit et al. 2011; Noll et al. 2014).

Being among the brighter transneptunian objects, Varda has been an attractive target for subsequent observational study. Fornasier et al. (2009) reported a featureless, reddish spectrum between 0.44 and 0.93 µm, with a slope of 19.2 ± 0.6 percent rise per 100 nm. Perna et al. (2010, 2013) used visible color photometry to classify the object as belonging to the "IR" spectral group of Barucci et al. (2005). A near-infrared spectrum presented by Barucci et al. (2011) showed no evidence of the $H_2O$ ice absorptions at 1.5 and 2 µm that appear in some TNO spectra, but a decline in albedo over the 2.05 to 2.3 µm wavelength range was tentatively attributed to absorption by methanol ice. Thirouin et al. (2010, 2014) presented photometric evidence for a low-amplitude single-peaked lightcurve with a period of 5.91 hours, although 4.76 and 7.87 hour periods were not formally excluded. For a double-peaked lightcurve, these would correspond to rotational periods of 11.82, 9.52, or 15.74 hours.

Varda's companion Ilmarë was discovered in 2009 at a separation of about 0.12 arcsec by K.S. Noll et al., using the Hubble Space Telescope (HST; Noll et al. 2011). Follow-up observations of the pair were done through a subsequent HST program, and additional near-infrared wavelength observations were obtained using laser guide star adaptive optics on ground-based telescopes Keck II and Gemini North, both located at the summit of Mauna Kea on the big island of Hawai'i. The next section describes, in chronological order, these observations and how they were processed.

## 2. Observational Data

The HST observation that discovered Varda's companion Ilmarë was part of Cycle 16 program 11113, led by K.S. Noll. It used the planetary camera (PC) of the Wide Field and Planetary Camera 2 (WFPC2; McMaster et al. 2008). The observing sequence (or "visit") consisted of four consecutive 260 second exposures using the *F606W* filter, a wide filter with a nominal effective wavelength of 606 nm.  The images were dithered by non-integer pixel offsets to improve spatial sampling of the scene, since at those wavelengths, the pixel scale of the WFPC2/PC (46 mas/pixel) under-samples HST's point spread function (PSF).  Thanks to the exceptionally stability of HST's PSF and the ability of the Tiny Tim software package to model it (e.g., Krist et al. 2011), this under-sampling is less of a problem than it otherwise would be.  For each image frame, we simultaneously fit a pair of Tiny Tim PSFs representing Varda and Ilmarë.  The scatter of the measurements from the four frames was used to estimate uncertainties on their relative positions and brightnesses.  Additional details of our WFPC2 data processing pipeline are provided by Grundy et al. (2009) and Benecchi et al. (2009).

Our next opportunity to obtain spatially resolved images of the system was with the Keck II telescope's NIRC2 near-infrared camera.  Under reasonable seeing conditions, the telescope's laser guide star adaptive optics (LGS AO) system enables NIRC2 to achieve spatial resolutions comparable to HST (e.g., Le Mignant et al. 2006).  To use LGS AO for faint targets like Varda and Ilmarë, a nearby (< 30 arcsec) and moderately bright ($R < 18$ mag) star is required as a reference for tip-tilt correction.  Our observations were done in $H$ band (1.49 to 1.78 µm), using stacks of three consecutive 100 second integrations followed by a dither, three more integrations, another dither, and then three final integrations.  Astrometric reduction of each stack of three frames was done by means of PSF fitting, using an azimuthally symmetric Lorentzian PSF.  Its width was fitted simultaneously with the positions of the two components of the binary.  We assumed a mean plate scale of 9.952 mas/pixel and an orientation offset of 0.252° (e.g., Konopacky et al. 2010; Yelda et al. 2010).  Based on experience from other binary observations with this system, we assumed a 1-σ uncertainty floor of ±2 mas on the relative astrometry.  A second, identical visit was done 1.8 hours later.  The observed change in relative positions between the two visits enabled us to determine that the sense of Ilmarë's orbital motion was clockwise on the sky plane, and also that the orbital period must be relatively short.  No photometric standards were observed, and no effort was made to compute $H$ band magnitudes from these data, which were taken solely for astrometric purposes.

Further HST observations of the system were obtained during four visits as part of Cycle 18 program 12237, led by W.M. Grundy.  By then, the fourth servicing mission to HST had replaced WFPC2 with the Wide Field Camera 3 (WFC3; Dressel 2015).  Each visit consisted of four dithered *F606W* images using the UVIS1 CCD detector.  Thanks to the greater efficiency of WFC3 compared with WFPC2, we were able to use almost as long integration times in the *F606W* images, and still have time left over for four additional images using another filter to obtain some spectral information.  To minimize effects of potential lightcurve variability on derived colors, these were sequenced with the *F606W* images being split into two groups of two, bracketing the other filter (except for the last visit where the final two *F606W* frames were lost).  The other filter was *F438W* in the first visit and *F814W* in the second visit (with nominal central wavelengths of 438 nm and 814 nm, respectively).  In the third and fourth visits, we used the infrared camera instead of UVIS1, obtaining two *F110W* images and two *F160W* images in each of those two visits, with central wavelengths 1.1 and 1.6 µm, respectively.  The infrared data will be presented in a separate paper.  As described in Grundy et al. (2012), PSF-fitting procedures for WFC3 data were similar to those used for WFPC2, with a pair of Tiny Tim PSFs being fitted to each frame independently.  Within each visit, the scatter in modeled relative positions and fluxes between frames was used to estimate the uncertainties in the combined measurements of those parameters for the visit.  The flexibility of the HST scheduling process allowed us to use optimal scheduling techniques (e.g., Grundy et al. 2008) to exploit the growing pool of astrometric information to inform scheduling of each successive visit.

A final series of observations was obtained with the Near-Infrared Imager (NIRI) camera at the Gemini North telescope (Hodapp et al. 2003), thanks to three-year NOAO survey program number 11A-0017, led by W.M. Grundy. Like the Keck II observations, the NIRI observations were done at near infrared *H* band wavelengths, with the use of a laser guide star with the Altair adaptive optics system (Herriot et al. 2000) being enabled by stellar appulses closer than 25 arcsec for stars brighter than $R < 16.5$. The Gemini observations were scheduled in queue mode, but the combination of NIRI with Altair and LGS was sparsely scheduled, so there was little scope for optimal scheduling. NIRI images were obtained in sets much like the Keck NIRC2 images, with a series of three to four exposures of 90 to 150 seconds in each of four dither positions. The larger/longer numbers were used for fainter targets, and for particularly faint targets, the entire pattern would be repeated a second time, but the Varda system being relatively bright for a Kuiper belt object, each observation of that system was done using the smaller numbers. The telescope was tracked at ephemeris rates. We used Lorentzian PSF profiles, but for some images had to include an ellipticity component where the PSFs were clearly elongated. The World Coordinate System (WCS) information in the image headers was used to convert from pixel coordinates to sky coordinates, via the xyad routine in the Astronomy User's Library (available from http://idlastro.gsfc.nasa.gov). We assumed a 1-σ uncertainty floor of ±3 mas for NIRI data. As with the Keck observations, we made no effort to do photometric calibration.

Example images from all four instruments are shown in Fig. 1, scaled to a common spatial scale and orientation. Also shown are the PSF models fitted to each image along with the residual images. The relative astrometry is compiled in Table 1.

The HST observations were processed to derive separate photometry of Varda and Ilmarë as shown in Table 2. The data reduction pipeline for PSF fitting photometry and conversion to standard Johnson *B*, *V*, and *I* magnitudes was described in detail in Benecchi et al. (2009 Appendix A; 2011). We used the zero-points provided by STScI in the image header to obtain the object magnitudes in the measured HST filters and then converted those values using `synphot` to get them in the standard Vegamag magnitude system. Finally we used the HST color information and an appropriately reddened Kurucz model solar spectrum convolved with the filter to convert the HST magnitudes to comparable standard ground based colors. On two of the HST visits, more than one filter was used, enabling us to compute color differences $B–V = 0.892 \pm 0.028$ and $V–I = 1.133 \pm 0.034$ for Varda and $B–V = 0.857 \pm 0.061$ and $V–I = 1.266 \pm 0.052$ for Ilmarë. These measurements show the two bodies to have *B–V* colors consistent with one another to within measurement

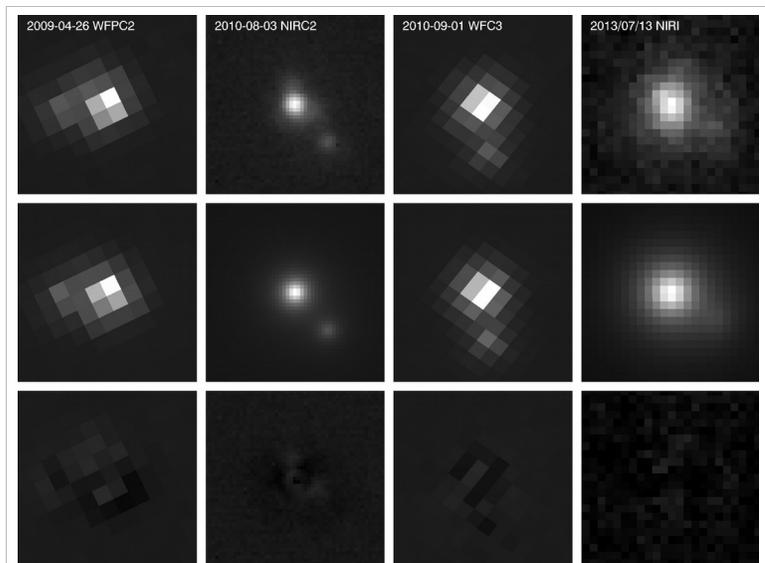

**Fig. 1.** Example images of Varda and Ilmarë from the four different instruments are shown across the top row, centered on Varda and projected to the sky plane with north up and east to the left. Each sub-panel is 0.5 arcsec across. The middle row shows PSF-fitted models of the images, and the bottom row shows the residuals. In each column, the stretch is consistent for all three images.

| Table 1. Observations of Relative Astrometry of Varda and Ilmarë | | | | | | |
|---|---|---|---|---|---|---|
| UT date and hour | Telescope/ instrument | r (AU) | Δ (AU) | g (deg.) | Δx (arcsec) | Δy (arcsec) |
| 2009/04/26  9$^h$.9311 | HST/WFPC2-PC | 47.842 | 47.055 | 0.76 | +0.1231(33) | −0.0103(20) |
| 2010/08/03  6$^h$.5746 | Keck II/NIRC2 | 47.687 | 47.215 | 1.09 | −0.0921(20) | −0.1086(20) |
| 2010/08/03  8$^h$.3464 | Keck II/NIRC2 | 47.687 | 47.216 | 1.09 | −0.0801(20) | −0.1115(20) |
| 2010/08/31  5$^h$.3457 | HST/WFC3-UVIS | 47.677 | 47.631 | 1.21 | −0.13093(97) | −0.0017(25) |
| 2010/09/01  13$^h$.0307 | HST/WFC3-UVIS | 47.677 | 47.652 | 1.21 | −0.0223(13) | −0.1332(14) |
| 2010/09/27  14$^h$.1742 | HST/WFC3-UVIS | 47.668 | 48.047 | 1.11 | −0.0056(47) | +0.1380(17) |
| 2011/07/09  18$^h$.9640 | HST/WFC3-UVIS | 47.571 | 46.795 | 0.80 | +0.0824(86) | −0.1117(41) |
| 2012/04/02  12$^h$.9388 | Gemini N./NIRI | 47.480 | 47.016 | 1.07 | −0.1131(30) | +0.0776(30) |
| 2012/05/06  11$^h$.2338 | Gemini N./NIRI | 47.469 | 46.626 | 0.68 | −0.0503(30) | +0.1307(30) |
| 2012/08/02  8$^h$.5709 | Gemini N./NIRI | 47.439 | 46.931 | 1.07 | −0.101(14) | −0.0778(62) |
| 2013/04/22  11$^h$.5694 | Gemini N./NIRI | 47.350 | 46.651 | 0.88 | −0.0840(30) | +0.098(11) |
| 2013/07/13  6$^h$.1370 | Gemini N./NIRI | 47.322 | 46.562 | 0.82 | −0.1220(33) | −0.0810(79) |

Table notes:

[a.] The distance from the Sun to the target is $r$ and from the observer to the target is $\Delta$. The phase angle $g$ is the angular separation between the observer and Sun as seen from the target.

[b.] Relative right ascension $\Delta x$ and relative declination $\Delta y$ are computed as $\Delta x = (\alpha_2 - \alpha_1)\cos(\delta_1)$ and $\Delta y = \delta_2 - \delta_1$, where $\alpha$ is right ascension, $\delta$ is declination, and subscripts 1 and 2 refer to Varda and Ilmarë, respectively. Estimated 1-$\sigma$ uncertainties in the final two digits are indicated in parentheses. Uncertainties are estimated from the scatter between fits to individual frames.

uncertainties, but their *V–I* colors could differ slightly, with Ilmarë's *V–I* color being redder than Varda's by 0.133 ± 0.062 mag. The possibility that their colors could be slightly different at these wavelengths has only 2-σ confidence and needs confirmation. Benecchi et al. (2009) showed that the colors of the components of most transneptunian binaries are correlated. A large color difference between binary components could provide a clue to formation via a different mechanism from that responsible for the more usual binaries with shared colors, or it could be related to a subsequent size-dependent process acting to alter the surfaces of the two bodies.

Our colors can also be compared with colors reported in the literature. Perna et al. (2010, 2013) reported several *B*, *V*, *R*, and *I* photometric measurements for the combined system from observations in 2008 and 2009. Weighted averages of their *B–V* and *V–I* colors are 0.893 ± 0.026 and 1.064 ± 0.031, respectively. These colors can be compared with merged colors from our observations, *B–V* = 0.886 ± 0.025 and *V–I* = 1.156 ± 0.029, computed by adding together

| Table 2. Separate CCD Photometry for Varda and Ilmarë | | | | | | | |
|---|---|---|---|---|---|---|---|
| UT Date | Varda | | | Ilmarë | | | $\Delta_{mag}$ |
| | *B* | *V* | *I* | *B* | *V* | *I* | |
| 2009/04/26 | - | 20.515(23) | - | - | 22.120(27) | - | 1.605(35) |
| 2010/08/31 | 21.170(21) | 20.278(18) | - | 22.944(34) | 22.087(51) | - | 1.786(32) |
| 2010/09/01 | - | 20.308(17) | 19.175(29) | - | 22.055(33) | 20.789(40) | 1.699(64) |
| 2010/09/27 | - | 20.244(17) | - | - | 21.837(70) | - | 1.593(72) |
| 2011/07/09 | - | 20.069(10) | - | - | 21.864(24) | - | 1.795(26) |

Table note

[a.] Photometric uncertainties were estimated from the scatter between multiple frames. Photometry was converted from HST filters *F438W*, *F606W*, and *F814W* into Johnson *B*, *V*, and *I* magnitudes using `synphot` as described in detail by Benecchi et al. (2009, 2011). Magnitude differences $\Delta_{mag}$ between primary and secondary are computed from all available filters used on each date.

our separate Varda and Ilmarë fluxes. The ground-based and HST colors agree reasonably well.

A parameter that will be needed for subsequent analysis is the relative brightness of Varda and Ilmarë, expressed as the difference between their brightnesses in magnitudes, $\Delta_{mag}$. We computed this for each of the HST observations shown in Table 2, and averaged them to obtain $\Delta_{mag}$ = 1.734 ± 0.042 mag. Although the *H* band Keck/NIRC2 images were not absolutely calibrated, we were able to measure the relative brightness of Varda and Ilmarë in them. Varda was brighter by 1.758 ± 0.044 magnitudes in that wavelength band (1.49 to 1.78 μm). The similarity to the visible wavelength $\Delta_{mag}$ suggests the two bodies share their common colors into the near-infrared.

Another useful parameter is the *V* band absolute magnitude $H_V$, which we can compute from our *V* photometry by removing the inverse distance squared effects of heliocentric and geocentric distances, *r* and *Δ*. The dependence on phase angle *g* differs from one object to the next and we do not know what value is appropriate for Varda and Ilmarë. If we assume no dependence on *g*, we obtain a weighted average $H_V$ = 3.246 ± 0.062 mags, whereas if we assume a more realistic *G* = 0.15 dependence in the *H* and *G* system (Bowell et al. 1989), we obtain $H_V$ = 3.097 ± 0.062 mags.

## 3. Mutual Orbit

Keplerian orbits were fitted to the relative astrometry in Table 1, accounting for the time-variable geometry between Earth and Varda. Procedures for finding the best-fitting orbit and assessing uncertainties of the orbital parameters have been described in several previous publications and the reader is referred to those papers for details (e.g., Grundy et al. 2011, and references therein). Briefly, Keplerian orbital elements were projected to the viewing geometry for the time of each observation and the $\chi^2$ goodness of fit computed. The orbital elements were iteratively adjusted to minimize $\chi^2$ by means of the Amoeba algorithm (Press et al. 2007). Uncertainties in the fitted parameters were estimated by adding Gaussian noise to the observed astrometry consistent with the reported error bars, and performing a new fit. This step was repeated many times to build up a Monte Carlo collection of fitted orbital elements, enabling uncertainties in the fitted parameters to be estimated from their dispersion.

As is often the case for trans-neptunian binaries, two viable orbit solutions exist, mirror images of one another through the mean sky plane over the course of the observations. The two solutions are compared with the observations in Fig. 2. This plot also shows that the observations sample orbital longitudes reasonably well, with only the North-East quadrant of the orbit having not been observed. Residuals for the two orbit solutions are shown in Fig. 3.

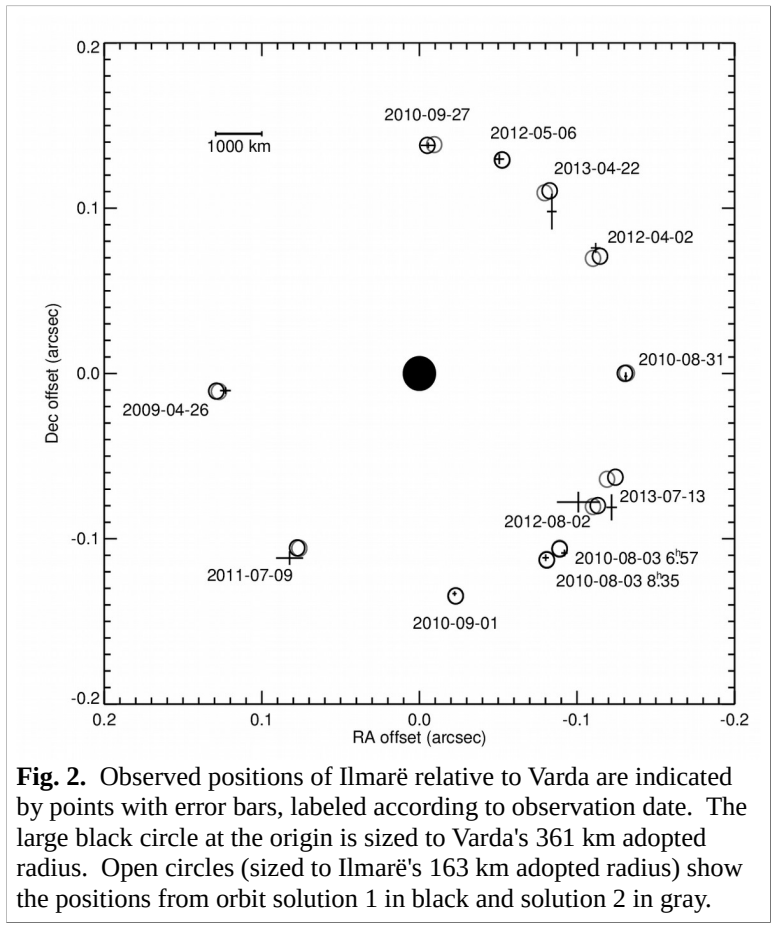

**Fig. 2.** Observed positions of Ilmarë relative to Varda are indicated by points with error bars, labeled according to observation date. The large black circle at the origin is sized to Varda's 361 km adopted radius. Open circles (sized to Ilmarë's 163 km adopted radius) show the positions from orbit solution 1 in black and solution 2 in gray.

The orbital parameters and $\chi^2$ goodness of fit corresponding to the two mirror solutions are listed in Table 3. Each observation provides two independent constraints ($\Delta x$ and $\Delta y$) so with twelve observations we have twenty four measurements constraining seven free parameters, leaving seventeen degrees of freedom. If the astrometric errors obeyed Gaussian distributions consistent with the stated uncertainties, and the observations could be considered truly independent of one another, we would expect $\chi^2 \approx 17$ for the true orbit. The larger $\chi^2$ values of our orbit solutions could be indicative of a slight underestimation of our astrometric error bars or they could point to inconsistent systematic errors associated with one or more of the four different imaging instruments used to observe the system. As a test of this possibility, we compared fits with and without the NIRI data. Those solutions were the same to within the estimated uncertainties on the fitted parameters.

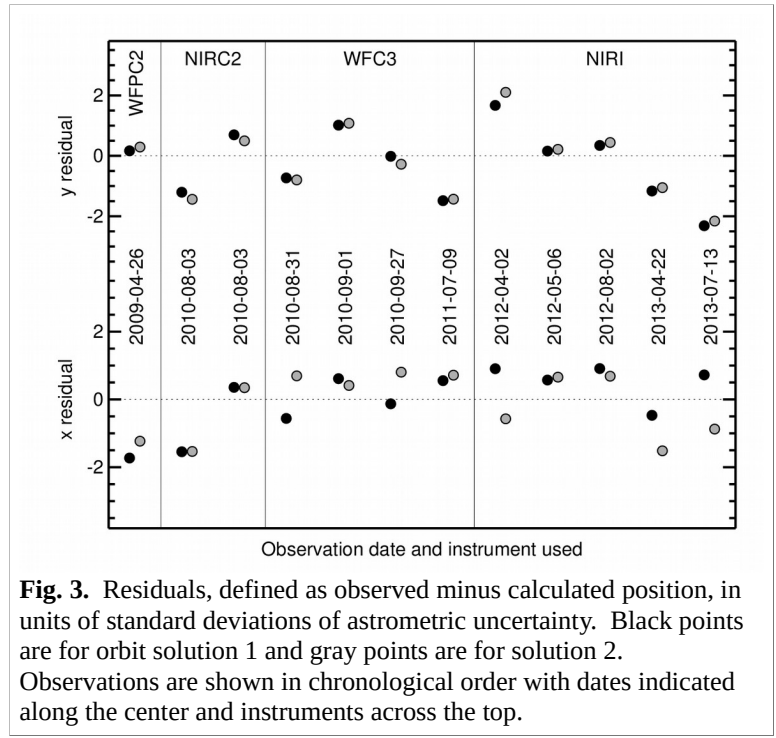

**Fig. 3.** Residuals, defined as observed minus calculated position, in units of standard deviations of astrometric uncertainty. Black points are for orbit solution 1 and gray points are for solution 2. Observations are shown in chronological order with dates indicated along the center and instruments across the top.

Table 3 also lists several parameters derived from the fitted orbital elements. These include the total mass of the system, the orientation of the pole of the mutual orbit, and the inclination between the mutual orbit and the heliocentric orbit. Twice during each 294 year heliocentric orbit, the plane of the mutual orbit sweeps across the inner Solar System. During these times, mutual events occur, with the two bodies occulting and/or eclipsing one another, providing additional opportunities for characterizing the system. For the two mirror orbit solutions, the next mutual event seasons would be centered on the years 2093 or 2064.

Eventually it will be possible to eliminate one of the two mirror solutions, thanks to the slow accumulation of parallax from Varda's orbital motion around the Sun. The evolution of the appearance of the two orbits is shown in Fig. 4. Already by 2020, sufficient parallax will have accumulated so that a strategically timed observation could easily break the ambiguity. In the mean time, it is useful to combine the two orbit solutions and adopt values for the period $P$, semimajor axis $a$, and eccentricity $e$ with associated uncertainties consistent with both orbit solutions. We do this for each parameter by adopting the mid-point between the extrema of its value plus and minus its 1-$\sigma$ uncertainty for both orbit solutions, resulting in the minimum symmetric error bar that encompasses both solutions and their 1-$\sigma$ uncertainties. These adopted values are listed in Table 4, and used in subsequent calculations through the remainder of the paper.

| Table 3. Varda – Ilmarë Mutual Orbit Solutions and 1-$\sigma$ Uncertainties | | | |
|---|---|---|---|
| Parameter and units | | Orbit 1 | Orbit 2 |
| **Fitted elements**[a] | | $\chi^2$ = 24.6 | $\chi^2$ = 26.9 |
| Period (days) | $P$ | 5.75063 ± 0.00010 | 5.75054 ± 0.00011 |
| Semimajor axis (km) | $a$ | 4812 ± 35 | 4805 ± 35 |
| Eccentricity | $e$ | 0.0181 ± 0.0045 | 0.0247 ± 0.0048 |
| Inclination[b] (deg) | $i$ | 101.0 ± 1.9 | 85.1 ± 1.8 |
| Mean longitude[b] at epoch[c] (deg) | $\epsilon$ | 95.6 ± 1.7 | 50.9 ± 1.5 |
| Longitude of ascending node[b] (deg) | $\Omega$ | 3.0 ± 1.5 | 319.5 ± 1.5 |
| Longitude of periapsis[b] (deg) | $\varpi$ | 304 ± 19 | 273 ± 13 |
| **Derived parameters** | | | |
| Standard gravitational parameter $GM_{sys}$ (km$^3$ day$^{-2}$) | $\mu$ | 17.82 ± 0.39 | 17.74 ± 0.39 |
| System mass[e] (10$^{18}$ kg) | $M_{sys}$ | 267.0 ± 5.8 | 265.8 ± 5.8 |
| Orbit pole right ascension[b] (deg) | $\alpha_{pole}$ | 273.0 ± 1.5 | 229.5 ± 1.5 |
| Orbit pole declination[b] (deg) | $\delta_{pole}$ | −11.0 ± 1.9 | 4.9 ± 1.8 |
| Orbit pole ecliptic longitude[d] (deg) | $\lambda_{pole}$ | 273.0 ± 1.5 | 225.7 ± 1.6 |
| Orbit pole ecliptic latitude[d] (deg) | $\beta_{pole}$ | 12.4 ± 1.8 | 22.3 ± 1.9 |
| Inclination between mutual and heliocentric orbits (deg) | | 99.1 ± 1.9 | 82.6 ± 1.9 |
| Next mutual events season | | 2093 | 2064 |

Table notes:
[a.] Elements are for Ilmarë relative to Varda. For Orbit solutions 1 and 2, $\chi^2$ is 24.6 and 26.9, respectively, based on observations at 12 epochs.
[b.] Referenced to J2000 equatorial frame.
[c.] The epoch is Julian date 2455300.0, corresponding to 2010-04-13 12:00 UT
[d.] Referenced to J2000 ecliptic frame.
[e.] Based on the CODATA 2006 value of the gravitational constant $G$ = 6.67428 × 10$^{-11}$ m$^3$ s$^{-2}$ kg$^{-1}$ (Mohr et al. 2008).

The system mass $M_{sys}$ derived from the mutual orbit is especially valuable if it can be combined with an independent estimate of the size of the bodies, in order to derive a density. Thermal emission from the Varda and Ilmarë system was observed by the European Space Agency's Herschel space observatory, as part of the "TNOs are Cool" open time key program (e.g., Müller et al. 2009). Using data from Herschel's Photodetector Array Camera and Spectrometer (PACS; Poglitsch et al. 2010), Vilenius et al. (2014) reported detecting the system at 70, 100, and 160 µm wavelengths. They used the fluxes in those three bands to estimate the combined surface area of the two bodies as being that of a single sphere of effective diameter $D_{eff} = 792^{+91}_{-84}$ km, which for our $H_V$ = 3.097 ± 0.060 corresponds to albedo $0.166^{+0.043}_{-0.033}$. Assuming Varda and Ilmarë have spherical shapes and both share the same albedo, we can use the brightness difference between them from our photometry ($\Delta_{\text{mag}}$ = 1.734 ± 0.042) to estimate their individual radii as $361^{+41}_{-38}$ and $163^{+19}_{-17}$ km for Varda and Ilmarë, respectively, with Varda accounting for 92% of the total volume. Further assuming both bodies have the same bulk density, the Varda/Ilmarë mass ratio would be the same as their volume ratio at $10.98^{+0.66}_{-0.62}$, only a little larger than the Pluto/Charon mass ratio of 8.58 (Tholen et al. 2008). The total mass can be divided by the total volume to get a bulk a density of $1.24^{+0.50}_{-0.35}$ g cm$^{-3}$. To obtain the asymmetric 1-σ uncertainty on the density we did a Monte Carlo calculation, randomizing $M_{\text{sys}}$, $D_{\text{eff}}$, and $\Delta_{\text{mag}}$ consistent with their uncertainties and combining them to assess the distribution of possible densities. Earlier, Vilenius et al. (2014) used a pre-publication version of our system mass to compute a density of $1.27^{+0.41}_{-0.44}$ g cm$^{-3}$, equivalent to our result within uncertainties, the differences being primarily attributable to our use of a slightly different $\Delta_{\text{mag}}$ value and a different method of estimating error bars.

| Table 4. Varda – Ilmarë Mutual Orbit Adopted Values and Derived Mass | | |
|---|---|---|
| Parameter and units | | Adopted value |
| Period (days) | $P$ | 5.75058 ± 0.00015 |
| Semimajor axis (km) | $a$ | 4809 ± 39 |
| Eccentricity | $e$ | 0.0215 ± 0.0080 |
| Gravitational parameter $GM_{sys}$ (km$^3$ day$^{-2}$) | $\mu$ | 17.78 ± 0.43 |
| System mass (10$^{18}$ kg) | $M_{sys}$ | 266.4 ± 6.4 |

Table notes:
[a.] Values and uncertainties are adopted for orbital parameters $P$, $a$, and $e$ so as to encompass both of the mirror orbit solutions. The system mass derived from these adopted values is used in subsequent computations.

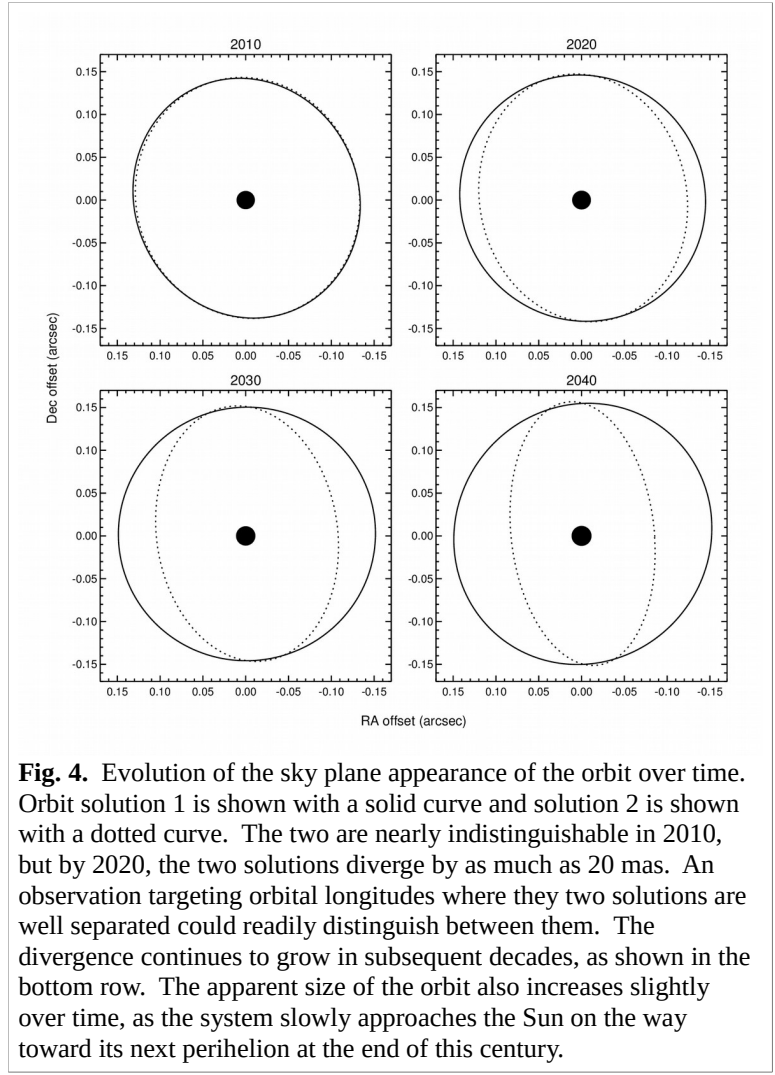

**Fig. 4.** Evolution of the sky plane appearance of the orbit over time. Orbit solution 1 is shown with a solid curve and solution 2 is shown with a dotted curve. The two are nearly indistinguishable in 2010, but by 2020, the two solutions diverge by as much as 20 mas. An observation targeting orbital longitudes where they two solutions are well separated could readily distinguish between them. The divergence continues to grow in subsequent decades, as shown in the bottom row. The apparent size of the orbit also increases slightly over time, as the system slowly approaches the Sun on the way toward its next perihelion at the end of this century.

We explored the sensitivity to our assumption of identical albedos for Varda and Ilmarë by postulating that one of the two objects has an albedo 50% greater than that of the other and redoing the computation. If Varda is the one with the higher albedo, its size must shrink relative to Ilmarë to maintain the observed flux ratio, resulting in radii of $347^{+40}_{-37}$ and $191^{+22}_{-21}$ km for Varda and Ilmarë, respectively. These radii give a smaller total volume for the system and thus a greater bulk density of $1.31^{+0.52}_{-0.36}$ g cm$^{-3}$. If Ilmarë has the higher albedo, the opposite happens, with the radii becoming $372^{+43}_{-39}$ and $137^{+16}_{-15}$ km and the bulk density dropping to $1.18^{+0.47}_{-0.33}$ g cm$^{-3}$. We did similar calculations to explore the effect of different densities. If instead of assuming both objects share the same density, we assume Ilmarë's bulk density is half that of Varda, the resulting bulk density for Varda then increases to $1.29^{+0.52}_{-0.36}$ g cm$^{-3}$ to absorb the mass transferred to it from Ilmarë. While the assumption of equal albedos and densities does influence the reported bulk density, relaxing these assumptions does not dramatically change it. The uncertainty in Varda's density is dominated by the uncertainty in volume.

Our bulk density for the Varda system is compared with densities of other transneptunian binaries in Fig. 5. In general, the smaller objects have low densities, mostly even lower than that of solid $H_2O$ ice, implying considerable internal void space. The largest four objects (Pluto, Eris, Quaoar, and Haumea) all have much higher densities, requiring a substantial contribution from materials denser than ice, presumably silicates. Varda's intermediate size and density place it in a transition zone between these two groups, accompanied by Salacia-Actaea, Orcus-Vanth, and 55637 2002 $UX_{25}$ (all individually labeled in Fig. 5). It could be tempting to draw conclusions about size or mass thresholds for onset of thermal differentiation but the picture is muddied by the fact that the objects in Fig. 5 are from diverse dynamical classes and thus likely formed at a variety of different heliocentric distances, potentially over different time scales and from different initial mixes of solids.

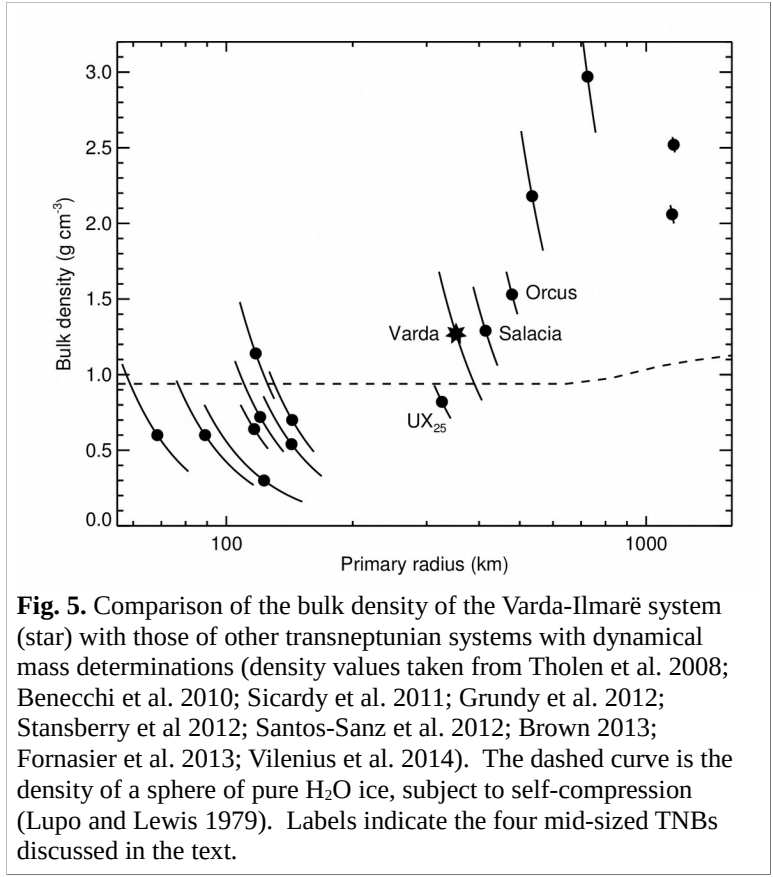

**Fig. 5.** Comparison of the bulk density of the Varda-Ilmarë system (star) with those of other transneptunian systems with dynamical mass determinations (density values taken from Tholen et al. 2008; Benecchi et al. 2010; Sicardy et al. 2011; Grundy et al. 2012; Stansberry et al 2012; Santos-Sanz et al. 2012; Brown 2013; Fornasier et al. 2013; Vilenius et al. 2014). The dashed curve is the density of a sphere of pure $H_2O$ ice, subject to self-compression (Lupo and Lewis 1979). Labels indicate the four mid-sized TNBs discussed in the text.

At Varda's perihelion distance of 39 AU, the system mass corresponds to a Hill radius of $2.1 \times 10^6$ km. The adopted Varda-Ilmarë semimajor axis of 4809 ± 39 km can be expressed as 0.23 percent of the Hill radius, making it the same as that of Salacia-Actaea (Grundy et al. 2011; Stansberry et al. 2012). By this measure, these two systems stand out as being the two most tightly bound of all transneptunian binaries with known orbits and masses. The orbital angular momentum $J_{orb}$ can be normalized for comparison to other TNB systems by dividing it by $J' = \sqrt{G M_{sys}^3 R_{eff}}$, where $R_{eff}$ is the radius of a single spherical body with the same total volume as the two components. For Varda and Ilmarë, $J_{orb}/J' = 0.28$. This specific orbital angular momentum is lower than those of the majority of TNBs with published orbits, although Salacia-Actaea, Orcus-Vanth, and 55637 2002 $UX_{25}$ have even lower values (e.g., Grundy et al. 2011; Brown 2013). Low values of $J_{orb}/J'$ have been interpreted as suggesting production of their satellites through low speed collisions or rotational fission rather than capture (e.g., Canup 2005; Descamps and Marchis 2008). However, it should be noted that Kozai cycles with tidal damping during periods of high eccentricity can also tighten TNB orbits from the initially looser orbits that result from various capture mechanisms (e.g., Perets and Naoz 2009; Porter and Grundy 2012). It may also be possible to directly co-accrete low specific angular momentum binary systems via gravitational collapse of nebular pebbles locally concentrated by streaming instabilities (e.g., Jansson and Johansen 2014), although the models of Nesvorný et al. (2010) tended to produce $J_{orb}/J'$ values of ~0.4 or greater. The mutual orbit of Varda and Ilmarë is highly inclined to its heliocentric orbit (see Table 3), another similarity with Salacia-Actaea, Orcus-Vanth, and 55637 2002 $UX_{25}$ (labeled in Fig. 5). For point masses, high inclinations would imply high amplitude Kozai cycles driven by solar tides, but Varda and Ilmarë are separated by only about 13 Varda radii so even very small deviations from spherical shapes would induce larger perturbations than solar tides would. The transition between wide, Kozai-dominated and tight, shape-dominated orbits is given by Nicholson et al. (2008, equation 3). Setting that criterion to the semimajor axis and solving for $J_2$ gives an extremely small oblateness limit ($J_2 \leq 2 \times 10^{-7}$) for solar tides to be competitive with perturbations from Varda's oblateness.

The mutual orbit of Varda and Ilmarë is circular, or nearly so. Our reported eccentricities for the two mirror orbit solutions are $0.0181 \pm 0.0045$ and $0.0247 \pm 0.0048$. The uncertainties on these values suggest 4-σ or 5-σ detections of non-zero eccentricities for the two solutions, respectively. A non-zero eccentricity is surprising because it is reasonable to expect the orbit to have become circularized via tidal interactions. A crude estimate of the time to circularize an initially eccentric orbit can be obtained by assuming the densities and sizes estimated earlier, along with dissipation parameter $Q = 100$, and rigidity $\mu = 4 \times 10^9$ N m$^{-2}$ (consistent with solid ice, e.g., Gladman et al. 1996; Grundy et al. 2007, 2011; Thirouin et al. 2014). The resulting time scale is around $10^8$ years, much less than the age of the Solar System, and it would be even more rapid for the smaller values of $\mu$ expected for more disrupted internal structures. Although the eccentricity of the Varda-Ilmarë orbit appears to be significantly non-zero, suggesting a relatively recent perturbation, we note that there are reasons for caution. Random astrometric data errors have a tendency to introduce eccentricity into orbit solutions, and in combining data from four different instruments, different systematic errors could enter. Additionally, our Monte Carlo method of estimating uncertainties could produce a skewed apparent probability distribution near zero eccentricity, since none of the Monte Carlo orbit solutions can have a negative eccentricity. Additional observations would be useful to confirm if the apparent small eccentricity is real. Fraser et al. (2013) reported a significant eccentricity in the mutual orbit of Quaoar and Weywot, another binary system that could reasonably have been expected to circularize. They discussed potential scenarios that could maintain or produce eccentricities in such a system.

Time scales to spin down the individual bodies can likewise be estimated as $10^8$ years to spin down Ilmarë and $10^{10}$ years to spin down Varda, from rapid initial rotations consistent with an oblique giant impact origin of the binary. Thus it would be reasonable to expect Varda's spin rate to not yet be synchronized to the mutual orbital period, while Ilmarë's spin should be locked, just as in the Earth-Moon system.

Thirouin et al. (2010, 2014) reported on photometric observations of the system carried out between 2006 and 2013. For detailed observational circumstances, see Thirouin et al. (2010) Table 1 and Thirouin et al. (2014) Table A.1. They used the Lomb (1976) method, as implemented by Press et al. (2007), commonly employed for searching for periodic signals in irregularly sampled time series data. From this analysis, Thirouin et al. reported low amplitude photometric variability with possible periods much shorter than the 5.75 day (138 hour) orbital period. The most probable period they found was 5.91 hours for a single-peaked lightcurve, corresponding to an 11.82 hour rotation period for an elongated body with a double-peaked lightcurve. Possible periods of 4.76 and 7.87 hours were also discussed, but considered less probable. If such short period photometric variability can be confirmed, it would mean that at least one of the bodies in the system is not tidally locked, presumably Varda. We merged the two Thirouin et al. data sets and recalculated their periodogram (see Thirouin et al. 2010 Fig. A.4 and Thirouin et al. 2014 Fig. A.1), focusing on the frequencies that would be expected for tidally locked components in the system, as shown in Fig. 6. A number of minor peaks appear, including a pair at 66.1 and 67.2 hours, highlighted by an arrow in the figure. These peaks are not far from the 69 hour period that would be expected for a tidally locked satellite with a double-peaked lightcurve, but their amplitude is low compared with the shorter period peaks discussed by Thirouin et al. and it would be inappropriate to ascribe significance to them. Even less power is seen at the 138 hour orbital period that would be expected for tidally locked bodies with single-peaked lightcurves like Pluto and Charon. More extensive photometric monitoring would be required to convincingly determine whether or not any of the bodies in the system has a spin rate consistent with the mutual orbital period. Assuming the spin poles coincide with the orbit pole, the objects are currently seen from high latitude, thus reducing the amplitude of any lightcurve variability they might have.

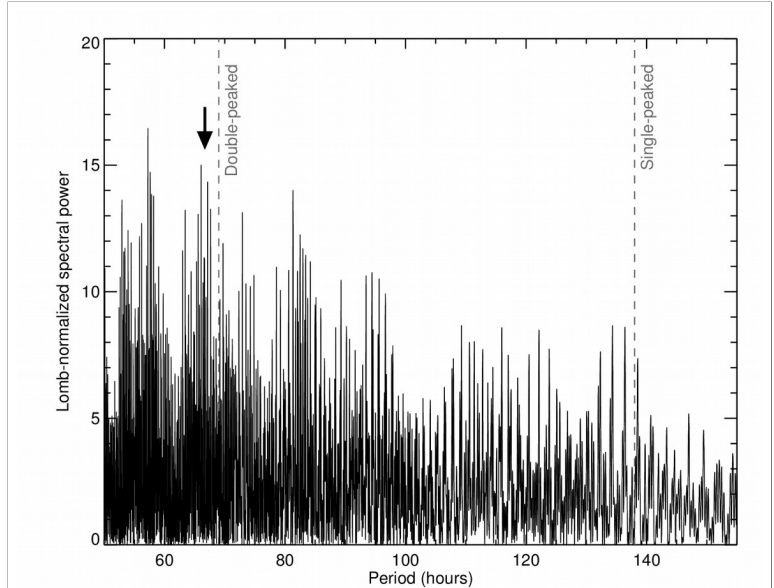

**Fig. 6.** Extension of the Thirouin et al. (2010 Figure A.4) periodogram to periods consistent with tidally locked spin rates, indicated by the two vertical dashed lines. The double-peaked lightcurve period would be expected for tidally-locked elongated shapes, while the single-peaked lightcurve period would be expected for a tidally-locked system with a lightcurve dominated by an albedo feature. Some peaks near 67 hours (indicated with an arrow) are close to, but do not exactly match the 69 hour period expected for a tidally locked body with a double-peaked lightcurve.

## 4. Summary and Conclusion

We report a sequence of visible and near-infrared images that spatially resolve Varda and Ilmarë, the primary and secondary components of a large transneptunian binary system with typical separations around 0.13 to 0.14 arcsec. Images were obtained at 12 distinct epochs spanning the interval from 2009 through 2013, using the Hubble Space Telescope and also laser guide star adaptive optics systems at Keck II and Gemini North telescopes on Mauna Kea. We measured the relative positions of Varda and Ilmarë at each epoch and used the data to derive their mutual orbit, with a period of 5.75 days and a semimajor axis of 4810 km. The orbit is nearly circular, and is seen nearly face-on. It is among the most tightly bound of all known transneptunian binary orbits, in terms of the semimajor axis as a fraction of the Hill radius. From the orbit we find a system mass of $2.7 \times 10^{20}$ kg, corresponding to roughly a tenth of the total mass of the asteroid belt (e.g., Petit et al. 2002, and references therein), but a much smaller fraction of the total mass of the Kuiper belt (Bernstein et al. 2004). In conjunction with a total surface area estimated from thermal radiometry (Vilenius et al. 2014), this mass gives a bulk density of 1.24 g cm$^{-3}$, intermediate between much lower densities found for small binary systems and the larger densities seen among the largest bodies in the Kuiper belt. The mutual orbit of Varda and Ilmarë is small compared with the Hill radius, has a small specific angular momentum compared with many TNBs, and is highly inclined with respect to the heliocentric orbit. These orbital characteristics are shared with other mid-sized TNBs Orcus-Vanth, Salacia-Actaea, and 55637 2002 UX$_{25}$, suggestive of a common formation and/or tidal-evolutionary history for these systems.

# Acknowledgments

This work is based in part on NASA/ESA Hubble Space Telescope programs 11113 and 12237. Support for these programs was provided by NASA through grants from the Space Telescope Science Institute (STScI), operated by the Association of Universities for Research in Astronomy, Inc., under NASA contract NAS 5-26555. We are especially grateful to Tony Roman at STScI for his efficiency in scheduling HST follow-up observations in program 12237.

Additional data were obtained at the W.M. Keck Observatory, which is operated as a scientific partnership among the California Institute of Technology, the University of California, and NASA and made possible by the generous financial support of the W.M. Keck Foundation. These data were obtained from telescope time allocated to NASA through the agency's scientific partnership with the California Institute of Technology and the University of California and their acquisition was supported by NASA Keck PI Data Awards, administered by the NASA Exoplanet Science Institute.

Additional data were obtained at the Gemini Observatory, operated by the Association of Universities for Research in Astronomy, Inc., under a cooperative agreement with the NSF on behalf of the Gemini partnership: the National Science Foundation (United States), the National Research Council (Canada), CONICYT (Chile), the Australian Research Council (Australia), Ministério da Ciência, Tecnologia e Inovação (Brazil) and Ministerio de Ciencia, Tecnología e Innovación Productiva (Argentina).

This work was supported in part by NSF Planetary Astronomy Grant AST-1109872. The authors wish to recognize and acknowledge the significant cultural role and reverence of the summit of Mauna Kea within the indigenous Hawaiian community. We are grateful to have been able to observe from this mountain. This manuscript benefited from reviews by D. Ragozzine and an anonymous reviewer. We thank them for the time and effort they contributed to improving the paper. Finally, we thank the free and open source software communities for empowering us with key tools used to complete this project, notably Linux, the GNU tools, LibreOffice, MariaDB, Evolution, Python, the Astronomy User's Library, and FVWM.